\documentclass[a4paper,conference]{IEEEtran}
\ifCLASSINFOpdf
   \usepackage[pdftex]{graphicx}
   \usepackage{epstopdf}
   \graphicspath{{../pdf/}{../jpeg/}}
   \DeclareGraphicsExtensions{.pdf,.jpeg,.png}

\else
   \usepackage[dvips]{graphicx}
   \graphicspath{{../eps/}}
\fi

\usepackage[cmex10]{amsmath}
\usepackage{amssymb}
\usepackage{amsthm}

\newcommand\blfootnote[1]{%
  \begingroup
  \renewcommand\thefootnote{}\footnote{#1}%
  \addtocounter{footnote}{-1}%
  \endgroup
}

\begin{document}
%
\title{Simultaneous Wireless Information and Power Transfer in a Two-User OFDM Interference Channel}
\author{\IEEEauthorblockN{Zati Bayani Zawawi*, Jaehyun Park${}^\dagger$ and Bruno Clerckx*${}^\ddagger$}
\IEEEauthorblockA{*Department of Electrical and Electronic Engineering, Imperial College London, London SW72AZ, United Kingdom.\\${}^\dagger$Department of Electrical Engineering, Pukyong National University, Korea.\\${}^\ddagger$School of Electrical Engineering, Korea University, Korea.\\
Email: z.zawawimohd-zawawi13@imperial.ac.uk, jaehyun@pknu.ac.kr, b.clerckx@imperial.ac.uk} \\[-6ex]}

\maketitle
\begin{abstract}
\boldmath
In this paper, we study the Simultaneous Wireless Information and Power Transfer (SWIPT) in a Single-Input Single-Output (SISO) two-user Orthogonal Frequency Division Multiplexing (OFDM) Interference Channel (IFC). We assume that the transmitters are non-cooperative and have perfect knowledge of the local Channel State Information (CSI). We show that the necessary condition for the optimal transmission strategy at high SNR is for the energy transmitter to transmit its signal by allocating its transmit power on a single subcarrier. Accordingly, we propose a one-subcarrier selection method for the energy transmitter and identify the achievable rate-energy region. In addition, we further enlarge the achievable rate-energy region by enabling a basic form of transmitter cooperation where messages are exchanged to inform the energy transmitter about the subcarriers unutilized by the information transmitter.
\end{abstract}
\vspace{-3pt}

\section{Introduction}
\vspace{-3pt}

Recently, Radio Frequency (RF) has been considered as a potential resource for energy harvesting in wireless system. Because an RF signal carries both information and power, SWIPT has been investigated [1]-[8]. There have been several studies on SWIPT addressing interference channel (IFC) \cite{Ref5}-\cite{Ref9}. SWIPT in a two-user and K-user MIMO system have been considered in \cite{Ref5} and \cite{Ref6}, respectively. \cite{Ref7} has considered SWIPT with partial CSI at the transmitter in MIMO IFC. SWIPT in a multi-user MISO IFC has been discussed in \cite{Ref8}. \cite{Ref9} has considered SWIPT in a two-user and K-user SISO IFC. Papers \cite{Ref1,Ref2} have discussed SWIPT in OFDM system but did not address the interference channel.

In this paper, we leverage our past results on SWIPT in MIMO IFC \cite{Ref5}-\cite{Ref7} to address SWIPT in SISO-OFDM IFC. To the best of the authors' knowledge, SWIPT for an OFDM interference channel has not been addressed yet\blfootnote{This work has been partially supported by the EPSRC of the UK under grant EP/M008193/1.}. We show that the optimal non-cooperative transmission strategy at the energy transmitter with local CSI knowledge at the transmitter (CSI of the links between a transmitter and all receivers) at high Signal-to-Noise Ratio (SNR) is to allocate its transmit power to a single subcarrier. In this paper, the optimality is at high SNR under the assumption that the transmitters are non-cooperative and have perfect knowledge of their local CSI. Accordingly, we propose a single subcarrier transmission startegy and identify the achievable rate-energy region. In addition, we show that by enabling some form of transmitter cooperation where messages are exchanged to inform the energy transmitter about the subcarriers unutilized by the information transmitter, we can further enlarge the achievable rate-energy region.

We note here that, the result of a one-subcarrier transmission strategy of an OFDM system can be regarded as the rank one beamforming of the MIMO system in \cite{Ref5}. However, the rank one beamforming of the MIMO system is in a continuous space domain while the subcarrier transmission strategy in the OFDM system is in a discrete frequency domain. This results in the selection of the optimal subcarrier and the computation of the achievable rate-energy (R-E) region being simpler than the design of the Geodesic beamformer and computation of the R-E region in MIMO \cite{Ref7}.

{\it Organization}: Section II presents the system model and section III discusses the transmission strategy.
Section IV presents the simulation results and section V concludes the paper. 

{\it Notations}: Bold capital letter and lower case letter represent matrix and vector, respectively. $(\bf A)^H$, $\mbox{tr}(\bf A)$, $\det(\bf A)$ and $\left\Vert{}\bf A\right\Vert{}$ represent the conjugate transpose, the trace, the determinant and the 2-norm of a matrix $\bf A$, respectively. $\left\vert{}a\right\vert{}$ denotes the absolute value of $a$ , $(a)^+ \triangleq \max(a,0)$ and $\bf A \succeq {\bf 0}$ indicates that matrix $\bf A$ is a positive semi-definite matrix. The matrix with $\mbox{diag}(a_1,…..,a_N)$ is a block diagonal matrix with elements $a_n$ and ${\bf I}_N$ is $N\times N$ identity matrix.
\vspace{-4.5pt}

\section{System Model}
In this section, we introduce the OFDM system model for the SISO two-user IFC where information and energy transmitters transmit their OFDM signal to their corresponding (information decoding or energy harvesting) receivers. Each OFDM symbol has N subcarriers and each transmitter and receiver has a single antenna. We assume that each transmitter has perfect knowledge of the local CSI, i.e. CSI of the links between itself and all receivers. The received signal at i-th receiver, ${\bf y}_i\in{}\mathbb{C}^{N \times 1}$ can be written as
\vspace{-6.5pt}
\begin{equation}
{\bf y}_i=\sum_{k=1}^2{{\bf H}_{ik}{\bf x}}_k+{\bf n}_i,
\end{equation}
where ${\bf n}_i\in{}\mathbb{C}^{N \times 1}$ is a complex white Gaussian noise vector with a covariance matrix ${\sigma{}}_n^2{\bf I}_N$. For simplicity, we assume that ${\sigma{}}_n^2=1$. In addition, we assume that the channel is slow fading, the transmitters are perfectly frequency synchronous and the cyclic prefix is long enough, such that the channel encompassing all subcarriers can be represented by a diagonal matrix structure as  ${\bf H}_{ik}$=diag($h_{ik,1},…..,h_{ik,N}$). Here, $h_{ik,n}$ is the channel coefficient for the n-th subcarrier between the k-th transmitter and the i-th receiver. The relative path loss of the cross link compared to the direct link is defined as $\delta_{ik} \in [0,1]$. The vector ${\bf x}_k\in{}\mathbb{C}^{N \times 1}$ is the transmit signal at the k-th transmitter with a transmit power constraint $P$, such that $E\left[{\left\Vert{}{\bf x}_k\right\Vert{}}^2\right]\leq{}P$
for $k=1,2$. In addition, $x_{k,n}$ is the transmit signal for the n-th subcarrier at the k-th transmitter with a transmit power $p_{k,n}=E\left[{\vert{}x_{k,n}\vert{}}^2\right]$.

When the receiver operates in the ID  mode, the achievable rate at the i-th ID receiver is given by
\begin{equation}
R_i=\log_2\det({\bf I}_N+{\bf H}_{ii}{\bf P}_i{\bf R}_{-i}^{-1}{\bf H}_{ii}^H)
\end{equation}
where ${\bf P}_i=\mbox{diag}(p_{i,1},…..,p_{i,N}$) is the power allocation at the i-th transmitter with $\mbox{tr}({\bf P}_i)\leq P$ and ${\bf R}_{-i}$ is the covariance matrix of the noise plus interference at the i-th receiver given
by ${\bf R}_{-i}={\bf I}_N+\sum_{\substack{k=1\\i\neq {k}}}^{2}{\bf H}_{ik}{\bf P}_k{\bf H}_{ik}^H$.

When the receiver operates in the EH mode, the total harvested energy at the i-th receiver (specifically, the total harvested energy of the baseband signal as in [1]-[8]) is given by
\vspace{-4.5pt}
\begin{equation}
E_i={\zeta{}}_iE\left[{\Vert{}{\bf y}_i\Vert{}}^2\right]={\zeta{}}_i\sum_{k=1}^2\mbox{tr}\left({\bf H}_{ik}{\bf P}_k{\bf H}_{ik}^H+{\bf I}_N\right).
\end{equation}
For simplicity, we assume that the energy harvesting efficiency constant
${\zeta{}}_i=1$ and the noise power is negligible compared to the energy harvested from the EH transmitter such that
\begin {equation}
E_i \approx \sum_{k=1}^2\mbox{tr}\left({\bf H}_{ik}{\bf P}_k{\bf H}_{ik}^H\right).
\end{equation}
\vspace{-10.5pt}

In this paper, we assume that the interference signal from other transmitters is not decodable \footnote{If it is decodable, the interference term can simply be set to 0.}. Accordingly, it degrades the achievable rate performance at ID user. In contrast, EH circuit can harvest energy from the interference signal. Therefore, the interference signal is beneficial to increase the total harvested energy at EH receiver.
\vspace{-4.5pt}

\section{Transmission Strategy}
In this section, we discuss the transmission strategy. There are three different scenarios, namely two-ID users, two-EH users and one-ID user and one-EH user. In this paper, our main focus is on one-ID user and one-EH user scenario.
\vspace{-6.5pt}

\subsection{Necessary Condition For The Optimal Transmission Strategy}
In the two-ID user scenario, the objective is to obtain the maximum achievable sum rate subject to the transmit power constraint at each transmitter. Since the two users decode data information, the total harvested energy is equal to zero. The problem in the two-ID user interference channel has been considered in several researches. For example, non-cooperative game theory has been developed to maximize the achievable sum rate with no CSI sharing among transmitters. Each user iteratively updates their power allocation using waterfilling algorithm until they reach a Nash Equilibrium \cite{Ref11}.

In the two-EH user scenario, the objective is to maximize the total harvested energy subject to the transmit power constraint at each transmitter. Since the two users harvest RF energy, the achievable rate is equal to zero. The optimal transmission strategy in two-EH user IFC is for the k-th EH transmitter to allocate its full transmit power to the subcarrier corresponding to the $\arg \underset{n=1,.,N}{\max}\{|h_{1k,n}|^2+ |h_{2k,n}|^2\}$.

Now, let us consider one-EH user and one-ID user scenario (with their corresponding transmitter respectively denoted as EH transmitter and ID transmitter) as in Fig. 1. Without loss of generality, we assume that the first receiver is in EH mode and the second receiver is in ID mode, (${EH}_1,{ID}_2$). The discussion is straightforwardly extended to (${EH}_2,{ID}_1$). The achievable rate-energy region is then given by
\begin{equation}
\begin{split}
C_{\left(R-E\right)}(P)\triangleq \{{\
R}\leq{}{\log}_2\det\left({\bf I}_N+{\bf H}_{22}{\bf P}_2({\bf R}_{-2})^{-1}{\bf H}_{22}^H\right),\\\sum_{k=1}^2\mbox{tr}\left({\bf H}_{1k}{\bf P}_k{\bf H}_{1k}^H\right)\geq{}{E},\ {\
\mbox{tr}\left({\bf P}_k\right)\leq{}P,\ {\bf P}}_k\succeq {\bf 0}, k=1,2\}.
\end{split}
\end{equation}
Power allocation ${\bf P}_1$ is coupled with power allocation ${\bf P}_2$ which makes the problem above
non-convex with respect to ${\bf P}_1$ and ${\bf P}_2$. The following proposition gives a necessary condition for the optimal transmission strategy for the EH transmitter at high SNR  \footnote{Note that in practice, wireless power transfer operates at high SNR [1]}.

\begin{figure}[!t]
\quad \quad \quad \quad
\includegraphics[width=2.8in]{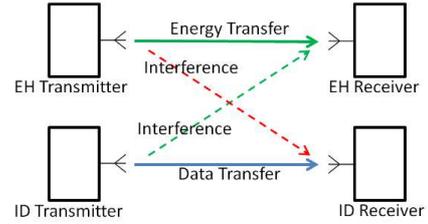}
\vspace{-6pt}
\caption{Two-user OFDM IFC in $(EH_1,ID_2)$ mode.}
\label{fig_sim}
\end{figure}

{\it Proposition 1:} The optimal transmission strategy at high SNR  for the EH transmitter is to transmit its signal by allocating its transmit power on a single subcarrier. 
\vspace{-7pt}

\begin{proof}
Let us consider the boundary point $({\bar R},{\bar E})$ of the achievable rate-energy, where $\mbox{tr}({\bf H}_{12}{\bf P}_2{\bf H}_{12}^H)\geq{\bar E}$. Then, for any given ${\bf P}_2$ on the boundary point, ${\bf P}_1={\bf 0}$ is the optimal power allocation at the EH transmitter. That is, because the harvested energy from the ID transmitter is sufficient to meet the energy constraint $\bar E$, EH transmitter does not transmit any signal causing the interference to the ID receiver. 

For $\mbox{tr}({\bf H}_{12}{\bf P}_2{\bf H}_{12}^H)<{\bar E}$, by letting ${\bf H}_{21}{\bf P}_1{\bf H}_{21}^H =\mbox{diag}(a_1,..,a_N)$ and ${\bf H}_{22}{\bf P}_2{\bf H}_{22}^H=\mbox{diag}(b_1,..,b_N)$, the boundary point of the achievable rate can be written as
\begin{equation}
{\bar R}={\log}_2\prod_{i=1}^N\left(1+\frac{b_i}{1+a_i}\right),
\end{equation}
with the harvested energy from the EH transmitter is given by $E_{11}=\mbox{tr}({\bf H}_{11}{\bf P}_1{\bf H}_{11}^H)={\bar E}-\mbox{tr}({\bf H}_{12}{\bf P}_2{\bf H}_{12}^H)$. Here, $b_i$ and $a_i$ is the received power from ID transmitter and the interference power from EH transmitter, respectively. Without loss of generality, let us first consider that the EH transmitter allocates its transmit power on $m$ subcarriers such that $m<N$ and the ID transmitter allocates its transmit power on $N$ subcarriers. For notation brevity, we sort the subcarrier indices at both transmitters such that the EH transmitter allocates its transmit power to the first $m$ subcarriers. Then, the achievable rate in (6) can be written as
\begin{equation}
{\bar R}=\log_2\Big(\prod_{i=1}^m\big(\frac{1+a_i+b_i}{1+a_i}\big)\prod_{j=m+1}^{N}\big(1+b_j\big)\Big).
\end{equation}
Because $b_i \propto P$ for $i=1,..,N$ and for a large $P$, $\log(1+b_i)\approx b_i$, (7) can be written as
\begin{equation}
{\bar R}\approx \log_2\Big(\prod_{i=1}^m\big(\frac{a_i+b_i}{1+a_i}\big)\prod_{j=m+1}^{N}b_j\Big).
\end{equation}
In addition, because $b_i \propto P$ and with a large $P$, $a_i$ is negligible with respect to $b_i$ when ${\bar E}$ is finite, then (8) can be written as
\begin{equation}
{\bar R}\approx \log_2\Big(\frac{\prod_{i=1}^Nb_i}{\prod_{i=1}^m(1+a_i)}\Big).
\end{equation}
That is, $\prod_{i=1}^m(1+a_i)$ in the denominator part of (9) has a minimum value when $m=1$ and $a_i$ (i.e., $a_1$) is small for given ${\bar E}$.
\end{proof}

Remark 1: Note that when ${\bar E}$ is large, both $b_i$ and $a_i$ are proportional to $P$. For large $P$, $\frac{a_i}{1+a_i} \approx 1$. Then, (8) can be written as
\begin{equation}
{\bar R} \approx \log_2\Big(\prod_{i=1}^m\big(1+\frac{b_i}{1+a_i}\big)\prod_{j=m+1}^{N}b_j\Big)
\end{equation}
where
\begin{equation}
{\bar R} \propto \log_2\Big(\frac{P^{N}}{P^m}\Big)=(N-m)\log_2P.
\end{equation}
That is, at high SNR and large ${\bar E}$, the achievable rate scales proportionally with $(N-m)$. Then, it is also maximized when $m=1$. 
\vspace{-4.5pt}


\subsection{Achievable Rate-Energy Region For One-ID User And One-EH User Scenario}
In this section, following the necessary condition of a single subcarrier transmission strategy at the EH  transmitter, we first discuss a single subcarrier selection strategies for the EH transmitter and accordingly we propose an iterative algorithm to identify the achievable rate-energy region. 

From an energy harvesting perspective, maximum harvested energy can be achieved when the EH transmitter allocates its transmit power to the subcarrier corresponding to the maximum channel gain (MaxCG) from the EH transmitter to the EH receiver. However, this may introduce a large interference to the ID receiver. From an information decoding perspective, the EH transmitter should allocate its transmit power to the subcarrier corresponding to the minimum channel gain (MinCG) from the EH transmitter to the ID receiver. However with MinCG, the EH receiver may not harvest sufficient energy to satisfy the energy constraint. In addition, motivated from \cite{Ref5} and \cite{Ref6}, the subcarrier index can be efficiently chosen such that the Signal to Leakage and Energy Ratio (SLER) is maximized, where SLER of the nth subcarrier is defined as
\begin{equation}
{\mbox{SLER}}_n=\frac{{\vert{}{h}_{11,n}\vert{}}^2}{{\vert{}{h}_{21,n}\vert{}}^2+\beta{}\left(\max\left(\left(\frac{\bar{E}}{P}-{\vert{}{h}_{11,n}\vert{}}^2\right),0\right)\right)}.
\end{equation}
Here, $\beta{}$ is a fixed regularization parameter. Therefore, the subcarrier index $\bar n$ is then chosen as
\begin {equation}
\begin{aligned}
&\arg \underset{n=1,.,N}{\max}{\vert{}{{h}_{11,n}\vert{}}^2} &&\mbox{for}\ \mbox{MaxCG} \\
&\arg \underset{n=1,.,N}{\min}{\vert{}{{h}_{21,n}\vert{}}^2} &&\mbox{for}\ \mbox{MinCG} \\
&\arg \underset{n=1,.,N}{\max}{\mbox{SLER}}_n &&\mbox{for}\ \mbox{SLER}
\end{aligned}
\end {equation}
\vspace{-9pt}

With the knowledge of the local CSI at the transmitter, we then have the following proposition.

{\it Proposition 2:} The optimal $\bar n$ for a single subcarrier transmission strategy at the EH transmitter is given by
\begin{equation}
\bar n=\arg \underset{n\in {\bf N}}{\max}\frac{{\vert{}{h_{11,n}}\vert{}}^2}{{\vert{}h_{21,n}\vert{}}^2},
\end{equation}
where ${\bf N}$ is a set of subcarrier indices at the EH transmitter satisfying the energy constraint in (5). Here, if ${\bf N}=\{\}$, then no feasible solution can satisfy the energy constraint in (5).
\vspace{-8pt}

\begin{proof}
This subcarrier selection strategy is an application of the Geodesic energy beamforming in (\cite{Ref7}, Proposition 3). Due to the limited space and the analogy with the proof of (\cite{Ref7}, Proposition 3); the proof is omitted.
\end{proof}
\vspace{-8pt}

Motivated by Proposition 2, we propose a new one-subcarrier selection method, namely Signal to Leakage Ratio with Energy Constraint (SLREC). $\bar{n}$ for SLREC is chosen as:
\begin{enumerate}
	\item According to the energy constraint in (5), possible subcarriers at the EH
transmitter are chosen such that $P{\vert{}{h_{11,n}}\vert{}}^2\geq\bar{E}-E_{12}$, where $E_{12}$ is the harvested energy from the ID transmitter.
	\item Among the possible subcarriers, $\bar n$ is the subcarrier index with the maximum ratio of
$\frac{{\left\vert{}h_{11,n}\right\vert{}}^2}{{\left\vert{}h_{21,n}\right\vert{}}^2}$.
\end{enumerate}

Note that in the SLREC, the searching process is required such that a set of subcarriers satisfying the energy constraint $\bar E$ is confined. In contrast, in the SLER, the searching process is not required. However, the subcarrier maximizing SLER is still implicitly considering $\bar E$ because the required energy $\bar E$ is included in the objective (12) of SLER.

Accordingly, we propose an iterative algorithm that optimizes the power allocation at the EH transmitter and the ID transmitter iteratively. We have the following optimization problem for the achievable rate-energy region of (5)
\begin {equation}
\begin{aligned}
(P1)\ &\mbox{maximize} &&\log_2\det\left({\bf I}_N+{\bf H}_{22}{\bf P}_2{\bf R}_{-2}^{-1}{\bf H}_{22}^H\right)\\
&\mbox{subject to} &&\sum_{k=1}^2\mbox{tr}\left({\bf H}_{1k}{\bf P}_k{\bf H}_{1k}^H\right)\geq{}\bar{E}\\
& &&\mbox{tr}({\bf P}_k)\leq P ,\ \ \ {\bf P}_k\succeq {\bf 0}\ \mbox{for}\ k=1,2, 
\end{aligned}
\end {equation}
where $\bar E$ is the energy constraint and ${\bf R}_{-2}={\bf I}_N + {\bf H}_{21}{\bf P}_1{\bf H}_{21}^H$. 

Because the objective function in optimization problem (P1) is monotonically decreasing with respect to ${\bf P}_1$, we optimize ${\bf P}_1$ using the steepest descent method. If the total harvested energy is larger than $\bar E$, EH transmitter reduces its transmit power to minimize the interference to the ID receiver. Given a matrix ${\bf A}(x)$,
$\frac{d}{dx}\ln{\det}{\bf A}\left(x\right)=
\mbox{tr}\left({\bf A}{(x)}^{-1}\frac{d{\bf A}(x)}{dx}\right)$. Then, the gradient of the achievable rate in (15) with respect to ${p}_{1,\bar n}$ is given by
\begin{equation}
\begin{split}
\bigtriangledown {\bf \it J}_{\bar{n}}({p}_{1,\bar{n}},{p}_{2,\bar{n}})&=\frac{1}{\ln(2)}\Big(
{(1+{p}_{1,\bar{n}}{\vert{}{h}_{21,\bar{n}}\vert{}}^2+{p}_{2,\bar{n}}{\vert{}{h}_{22,\bar{n}}\vert{}}^2)}^{-1}\\&-{(1+{p}_{1,\bar{n}}{\vert{}{h}_{21,\bar{n}}\vert}^2)}^{-1}\Big){\vert{}{h}_{21,\bar{n}}\vert{}}^2.
\end{split}
\end{equation}
\vspace{-10pt}

Given ${\bf P}_1$, the optimization problem (P1) is convex with respect to ${\bf P}_2$. Therefore, ${\bf P}_2$ at the i-th iteration can be solved using the Lagrangian duality method. Given ${\bf P}_1$, the Lagrangian of optimization problem (P1) is given by
\begin{equation}
\begin{split}
\mathcal{L}({\bf P}_2, \lambda,\mu)={\log}_2\det\left({\bf I}_N+{\bf H}_{22}{\bf P}_2({\bf R}_{-2})^{-1}{\bf H}_{22}^H\right)-\\
\mu\left(\mbox{tr}{({\bf P}_2)-P}\right)+
\lambda\left(\mbox{tr}\left({\bf H}_{12}{\bf P}_2{\bf H}_{12}^H-(\bar E-E_{11})\right)\right),
\end{split}
\end{equation}
where $\mu{}$ and $\lambda{}$ are the Lagrangian multipliers for the transmit power constraint and the energy constraint, respectively. ${\bf P}_2$ can be obtained by setting $\frac{d \mathcal{L}({\bf P}_2,\lambda,\mu)}{d{\bf P}_2}={\bf 0}$, such that
\begin{equation}
{\bf P}_2={\left(\gamma{}-{{\bf R}_{-2}}{({\bf H}_{22}{\bf H}_{22}^H)}^{-1}\right)}^+,
\end{equation}
where $\gamma{}={\big(\ln(2)(\mu{}I_N-\lambda{}{\bf H}_{12}{\bf H}_{12}^H)\big)}^{-1}$. Here, $\mu$ and $\lambda$ can be solved using the subgradient-based method \cite{Ref10}, where the subgradient of $g(\lambda,\mu)$ is given by $(\mbox{tr}\left({\bf H}_{12}{\bf P}_2{\bf H}_{12}^H\right)-(\bar E-E_{11}), P-\mbox{tr}({\bf P}_2))$. Note that if the harvested energy from the ID transmitter is sufficient, $\lambda{}=0$. Then, ${\bf P}_2$ is solved using a single-user waterfilling algorithm, maximizing the achievable rate with transmit power constraint.

The iterative algorithm to identify the achievable rate-energy (R-E) region is given in Algorithm 1.\\
Algo. 1. \underline{Identification of the R-E region}
\begin{enumerate}
	\item Initialize ${\bf P}_1^{(0)}$ such that ${p}_{1,\bar n}^{(0)}=P$ and ${p}_{1,n\not= \bar n}^{(0)}=0$. 
	\item For $i=0:I_{max}$, where $I_{max}$ is the maximum number of iterations.
	\begin{itemize}
	 \item[a)] For given ${\bf R}_{-2}^{(i)}$, update ${\bf P}_2^{(i)}$ as in (18).
	 \item[b)] If $\sum_{k=1}^2\mbox{tr}({\bf H}_{1k}{\bf P}_k^{(i)}{\bf H}_{1k}^H)>\bar{E}$\\
 	 Update $\bigtriangledown {\bf \it J}_{\bar{n}}$ as in (16). If $\bigtriangledown {\bf \it J}_{\bar{n}}=0,\ $then set $\bigtriangledown {\bf \it J}_{{\bar n}}=\alpha{}$, where $\alpha$ is a fixed negative parameter.
	Update ${\bf P}_{1}^{(i+1)}$  such that\[{p}_{1,\bar{n}}^{(i+1)}=\min(\max{\left({p}_{1,\bar{n}}^{(i)}+\Delta{}\bigtriangledown{\bf\it J}_{\bar{n}},0\right)},P).\]
	\end{itemize}
\item Finally, the achievable rate-energy is determined as $(R_2,E_1)=\Big({\log}_2\det({\bf I}_N+{\bf H}_{22}{\bf P}_2^{(i)}({\bf R}_{-2}^{(i)})^{-1}{\bf H}_{22}^H), \sum_{k=1}^2\mbox{tr}({\bf H}_{1k}{\bf P}_k^{(i)}{\bf H}_{1k}^H)\Big)$.
\end{enumerate}
The step size ${\Delta}$ is given by a fixed value in $[0,{\Delta{}}_{max}]$. The maximum allowable step size, ${\Delta{}}_{max}$ is given by \cite{Ref6}
\begin{equation}
\Delta_{max}=\frac{{\bar {E}}-\mbox{tr}({\bf H}_{11}{\bf P}_1^{(i)}{\bf H}_{11}^H)-\mbox{tr}({\bf H}_{12}{\bf P}_2^{(i)}{\bf H}_{12}^H)}{{\bigtriangledown {\bf \it J}_{\bar{n}}({p}_{1,\bar{n}}^{(i)},{p}_{2,\bar{n}}^{(i)}){\vert{}{h}_{11,\bar{n}}\vert{}}^2}}.\end{equation}

Due to the discrete subcarrier transmission strategy in OFDM system, when EH transmitter transmits its signal at the unutilized subcarrier of the ID user, the interference caused by the EH transmitter does not create interference at the ID receiver. Therefore in step 2. b) of Algorithm 1, when the ID transmitter does not allocate its transmit power at subcarrier $\bar{n}$ due to its channel condition, then $\bigtriangledown{\bf\it J}_{\bar{n}}=0$. Accordingly, the EH transmitter will decrease its transmit power at subcarrier $\bar n$ by setting $\bigtriangledown{\bf\it J}_{\bar{n}}=\alpha$. 

Algorithm 1 always has negative gradient and positive step size, such that ${\bf P}_{1}$ is monotonically decreasing. The objective function in the optimization problem (P1) is concave with respect to ${\bf P}_{2}$ and monotonically decreasing with respect to ${\bf P}_{1}$. That is, the objective function of (P1) is quasiconcave and the constraints are convex. Therefore, we can conclude that Algorithm 1 always converges to a global optimal solution for given local CSIT and subcarrier selection strategy \cite{Ref5}.

In the simulation results in section IV, an upper-bound on the R-E region achievable with the non-cooperative strategy with local CSIT and a single subcarrier selection strategy is obtained by conducting an Exhaustive Search over all possible subcarriers. Specifically, the Exhaustive Search algorithm performs a full search for the optimal single-subcarrier selection and power allocation for the EH transmitter, in which the ID transmission strategy (i.e. ID transmitter's power allocation) is known at the energy transmitter. For a given energy constraint, we evaluate the achievable rate from Algorithm 1 for each subcarriers at the EH transmitter. Finally, the EH transmitter allocates its transmit power at the subcarrier with the highest achievable rate. Note that Exhaustive Search algorithm requires transmitter cooperation with global CSIT knowledge (CSI of the links between all transmitters and all receivers).

Remark 2: In a single-user waterfilling algorithm, the ID transmitter does not allocate its transmit power to the subcarrier with poor channel gain-to-noise ratio depending on the ${\it water\ level}$. That is, if there is some interference on the subcarriers unutilized by the ID transceiver, it will not degrade the achievable rate performance. Hence, $\bigtriangledown {\bf \it J}_{\bar{n}}={\bf 0}$. However, for the EH transmitter to make use of the knowledge of those subcarriers unutilized by the ID transmitter, the EH transmitter requires some transmitter cooperation (as in the Exhausive Search algorithm) where the indices of those unutilized subcarriers are informed to the EH transmitter.
\vspace{-8.5pt}

\subsection{Information Sharing of Unutilized Subcarrier Indices}
\vspace{-2pt}

From Remark 2, by allowing the ID transmitter to share with the EH transmitter the indices of the subcarriers unutilized by the ID transmitter, we can further enlarge the achievable R-E region. Let ${\bf Z}$ be the set of unutilized subcarrier indices at the ID-user, i.e., ${\bf Z} =\{n_u| P_{2,n_u}=0\}$. $R$ in (5) can be written as
\begin{equation}
R = \sum_{n\notin {\bf Z}}^N \log_2\left(1+p_{2,n}\vert{}h_{22,n}\vert{}^2{(1+p_{1,n}\vert{}h_{21,n}\vert{}^2)}^{-1}\right).
\end{equation}
That is, the achievable rate is not degraded by the interference signal on the subcarrier unutilized by the ID transceiver. Therefore, if the EH transmitter knows the unutilized subcarrier indices (i.e., {\bf Z}), it can transfer the energy without interfering the ID operation. Accordingly, with a transfer of information from the ID transmitter to the EH transmitter, we can allocate the power to the subcarrier unutilized at the ID transceiver. The subcarrier index $\bar n$ for SLREC with information sharing is chosen as: 
According to the energy constraint in (5), possible subcarriers at the EH transmitter are chosen. Among them,
\begin{enumerate}
\item If there are unutilized subcarriers at the ID transmitter, choose $\bar n$ as the subcarrier index with the largest channel gain ${\vert{}{h}_{11,n}\vert{}}^2$.
\item Else, choose $\bar n$ among them as the subcarrier index with the maximum ratio of
$\frac{{\left\vert{}h_{11,n}\right\vert{}}^2}{{\left\vert{}h_{21,n}\right\vert{}}^2}$ (i. e., SLREC).
\end{enumerate}
The iterative algorithm to identify the achievable rate-energy region for SLREC with information sharing of unutilized subcarrier indices at ID transmitter is given in Algorithm 1.
\vspace{-5pt}

\section{Simulation Results}
\vspace{-3pt}

In this section, we present the simulation results. We assume that the path loss is set to $10^{-3/2}$ such that the channel ${\bf H}_{ik}=10^{-3/2} \sqrt{\delta_{ik}} \bar{\bf H}_{ik}$, where $\bar{\bf H}_{ik}=\mbox{diag}({\bf{F}}[\bar{\bf h}_{ik} \ {\bf 0}_{N-L}]^T)$. $\bar{\bf h}_{ik}\in{}\mathbb{C}^{1\times L}$ is the multipath channel of length $L=3$ where the frequency selective fading channel is modeled using three-tap exponentially distributed power profile, each with complex zero-mean random Gaussian distribution. ${\bf F}\in{}\mathbb{C}^{N \times N}$ is the normalized Fourier matrix. Here, $\delta_{ik}=1$ for $i=k$ and $\delta_{ik} $ $=0.8$ for $i\neq k$. In addition, $P$ is set to $50mW$, the noise power at each subcarrier is set to $1\mu$W and $N$ is set to $8$.

Fig. 2 shows the R-E tradeoff in the two-user OFDM IFC. The EH transmitter allocates its transmit power on a single subcarrier based on MaxCG, MinCG, SLER and SLREC as described in section III-B and the achievable rate-energy region is determined based on Algorithm 1. The simulation result shows that the MinCG increases the achievable rate at the ID receiver while the MaxCG achieves a larger harvested energy at the EH receiver. As expected, the R-E region for the SLER covers both MaxCG and MinCG R-E regions. The proposed SLREC subcarrier selection method has a larger R-E region than SLER. That is, SLREC minimizes the interference gain at ID receiver provided that the energy constraint is satisfied. In addition, in the region where the harvested energy is less than $18 \mu$W, the achievable rate is unchanged. Here, because the harvested energy from the ID transmitter is sufficient to meet the energy constraint, EH transmitter does not transmit any signal to prevent any interference to ID receiver.

Fig. 3 shows that SLREC with information sharing of the unutilized subcarriers at the ID transmitter has a larger achievable R-E region than SLREC without information sharing. That is, because the interference signal at the unutilized subcarriers at ID-user does not degrade the achievable rate, the EH transmitter can transfer the energy without compromising the achievable rate at the ID receiver. In addition, the Exhaustive Search algorithm provides an upper-bound on the achievable R-E region  with a single subcarrier selection strategy at the EH transmitter.
\begin{figure}[!t]
\centering
\includegraphics[width=2.05in]{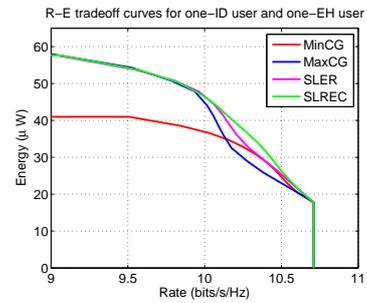}
\vspace{-10.5pt}
\caption{R-E tradeoff for MaxCG, MinCG, SLER and SLREC.}
\label{fig_sim}
\end{figure}
\begin{figure}[!t]
\centering
\includegraphics[width=2.05in]{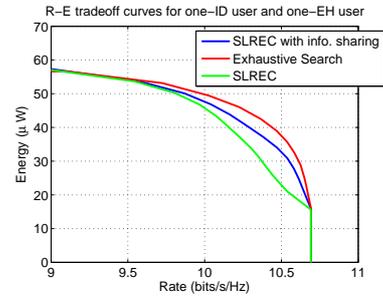}
\vspace{-10.5pt}
\caption{R-E tradeoff for SLREC, Exhaustive Search and SLREC with information sharing by ID-user.}
\label{fig_sim}
\end{figure}
\vspace{-9.5pt}

\section{Conclusion}
In this paper, we have discussed SWIPT in SISO two-user OFDM non-cooperative interference channel with perfect knowledge at the transmitters of the (local) CSI of the link between itself and all receivers. We have found that, the necessary condition for the optimal transmission strategy in high SNR is for the EH transmitter to allocate its transmit power on a single subcarrier. Accordingly, we identify the achievable rate-energy regions for different subcarrier selection strategies - MaxCG, MinCG, SLER and SLREC. We have found that SLREC exhibits higher R-E performance than the other subcarrier selection strategies. Interestingly, with some transmitter cooperation allowing the indices of the subcarriers unutilized by the ID transmitter to be shared with the EH transmitter, we can enlarge the achievable rate-energy region.

\end{document}